\documentclass[final,3p,times,twocolumn]{elsarticle}

\usepackage{amssymb}

\biboptions{sort&compress}

\usepackage[figuresright]{rotating}
\usepackage{mathtools}

\usepackage{multirow}
\usepackage{siunitx}

\usepackage{paralist}
\usepackage{subcaption}
\usepackage[dvipsnames]{xcolor}

\DeclarePairedDelimiterXPP\BigOSI[2]%
  {\mathcal{O}}{(}{)}{}%
  {\SI{#1}{#2}}

\begin{document}

\begin{frontmatter}

\title{Wide band spectroscopic response of monocrystallines to low dose neutron and gamma radiation}

\author[1]{Yossi Mosbacher {*}}
\author[1]{Micha Weiss}
\author[1]{Hagar Landsman}
\author[1]{Nadav Priel}
\author[2]{Ilan Eliyahu}
\author[2]{Arik Kreisel}
\author[2]{Offir Ozeri}
\author[2]{David Hershkovich}
\author[3]{Ori Cheshnovsky}
\author[1]{Ranny Budnik}

\cortext[*]{yossi.mosbacher@weizmann.ac.il}

\address[1]{Department of Particle Physics and Astrophysics, Weizmann Institute of Science, Rehovot, Israel}
\address[2]{Soreq Nuclear Research Center, Yavne 81800, Israel}
\address[3]{Raymond and Beverly Sackler School of Chemistry, Tel-Aviv University, Tel-Aviv, Israel}

\begin{abstract}
We identify a number of crystalline structures with promising characteristics to serve as a detection medium for a novel Dark Matter (DM) detector with a low threshold energy. A detector of this kind can be specifically useful in application requiring the detection of nuclear recoils, such as in direct detection of low mass DM, coherent neutrino scattering and neutrons. We describe a broad band, high sensitivity optical setup designed and constructed for the purpose of this search and future investigations of specific crystals. We report on the fluorescent signals produced from exposure to low doses of neutrons and $\gamma$ rays and find potential targets in Quartz, Sapphire, LiF, CaF$_{2}$ and BaF$_{2}$. These crystals and specific signals will be the subject of further study to establish the various traits relevant for a full scale DM detector. In this paper we identify the most interesting signals that will be promoted to significantly more detailed studies, including their production mechanism.
\end{abstract}


\end{frontmatter}

\section{Introduction}

\label{sec:intro}
The current understanding of the Universe  consists of the $\Lambda$CDM model, and requires that most of the matter in it be dark, i.e. almost non-interacting with standard model particles~\cite{Aghanim:2018eyx}. 
The most widely studied form of this postulated Dark Matter (DM) is the Weakly Interacting Massive Particle (WIMP), proposed in~\cite{Goodman:1984dc}.

Direct detection~\cite{Drukier:1986tm} experiments have focused over the last few decades on searches for WIMPs with masses
above $\sim$10~GeV/c$^2$, the emphasis being on detecting Nuclear Recoils (NRs) induced by WIMPs from the Milky-Way DM halo. These experiments are sensitive to recoils with energy deposition of more than $\BigOSI{}{keV}$, which sets the $\sim$10~GeV/c$^2$ DM mass threshold. 
The signal scattering off nuclei in various target materials must be separated from Electronic
Recoils (ERs) which constitute the major background in such experiments. 
So far no  signal  was identified, and as stronger experimental constraints are being set, more theoretical models are being challenged, 
 see~\cite{Roszkowski:2017nbc} for a review.  The experiments leading this field feature large exposures of order tonne$\cdot$year, and almost negligible background leaking into the signal region~\cite{Aprile:2018dbl}.

In recent years, a growing number of theoretical models suggest the existence of light DM below the GeV/c$^2$ scale~\cite{essig2010origin,boehm2004scalar,feng2008dark,hochberg2015model,an2015direct,hooper2008natural,d2015light,choi2012axino,falkowski2011asymmetric,lin2012symmetric,boddy2014self}. In these cases the expected energy deposited by the recoiling nucleus is lowered to the $\sim$eV~scale. As an example, a 100~MeV/c$^2$ DM particle  maximal recoil energy is  $E_{NR}\sim$10~eV  using standard DM halo parameters. 
This motivates expanding the direct detection technologies to allow lower energy thresholds,
often below the abilities of current direct detection technologies~\cite{Alexander:2016aln}.

The sensitivity for detection of DM particles lighter than a few GeV/c$^2$ has been strongly limited by the inherent challenges of lowering the detection threshold, elimination of backgrounds and the use of large exposures.

The lowest NR threshold today is set by cryogenic calorimeters, with
CRESST  achieving the most sensitive limit 

below GeV/c$^2$,  with an energy threshold of $\sim$100~eV~\cite{Petricca:2017zdp}, a 24~g target and a significant amount of background events.

In this paper we continue the work on a proposed detection method~\cite{Budnik:2017sbu}, following a general idea from~\cite{essig:2011nj, Essig:2016crl}. The method includes the use Color Centers (CCs) as a detection channel which has the potential to extend the sensitivity for DM detection down to masses of a few hundreds of MeV/c$^2$ offering scalability to large exposures. CCs are defects in normally transparent crystals, which, by the deformation of the periodic symmetry, create electronic states within the band gap~\cite{CCBook,Reuven}. These defects have various documented formation mechanisms \cite{ashcroft1976solid}, including NRs which cause dislocations of crystal nuclei. Electronic states of the CC, when occupied by electrons, can absorb optical photons, giving rise to coloration, from which the name is derived. These electrons are excited to a higher state, and can decay back by a combination of phonon emission and a fluorescence photon, at a longer wavelength compared to the exciting one.

Many of their properties make CCs ideal for the detection of a single dislocation event in large volume;
CCs live practically indefinitely at room temperatures and can be detected through their excitation and emission of luminescence. Formation of CCs in crystals typically requires atomic scale energies of $\BigOSI{10}{eV}$, orders of magnitude below current WIMP detection capabilities. A light DM particle recoiling off a nucleus inside a crystal can create an active CC that can be probed continuously through its characteristic fluorescence. Beyond its usefulness for DM searches, this channel is also sensitive to solar neutrino coherent scattering off the nucleus~\cite{Akimov:2017ade, Budnik:2017sbu}.

\subsection*{Experimental approach}
The path to the realization of a detector such as the one described above can be separated in to three parts;
\begin{inparaenum}[(i)] \item Identification of the most promising CC candidates, \item the investigation of their response to ER and NR on a single site level, and \item finally building a small scale prototype DM detector. 
\end{inparaenum}
In the first part, the results for which are reported in this work, 10 optically transparent and comercially available crystals, were irradiated by $\gamma$ and n sources, and measured to detect Radiation Induced Luminescence (RIL) in their absorption and emission spectra. While extensive work has been done on the effects of high doses of radiation on various crystals, e.g.~\cite{PhysRevB.9.775, IZERROUKEN2007696, GEE1980577, Abdukadyrova2007,allen1995point,PhysRev.186.926,CCBook}. This work is a methodical investigation of low doses aimed specifically at identifying potential discrimination between NRs and ERs in RIL formation.

In order to systematically study defect formation by scattering off a nucleus, NRs with the relevant momentum transfers are needed. This can be achieved using a controlled neutron source with an adequate energy range.
For DM, well below 1~GeV/c$^2$, comparable interactions can be produced by neutrons of $\BigOSI{10}{keV}$. Such low energy neutrons are difficult to produce with good energy resolution, at high rates in a controlled manner. 
Due to the low availability of $\BigOSI{10}{keV}$ neutron sources, it is necessary to first identify the most promising crystals, using a readily available $\BigOSI{}{MeV}$ neutron source. Neutrons from such a source provide signatures that partially overlap with those of lower energy neutrons~\cite{Mannhart1987}.\par
The discrimination factor is a very important figure of merit for DM detectors; it is the ability to differentiate the signal of a NR from ERs which are much more abundant in a typical experiment. For this reason, the crystals this study aims to identify must exhibit detectable NR induced luminescence and should be prioritized by their discrimination factor. The irradiations should also be confined to a relatively low dose regime, thus limiting interaction between defects and maintaining a linear response to the radiation dose~\cite{RabinKlik}. For this purpose, a $^{60}$Co $\gamma$ source is used to measure the effects of low doses of $\gamma$ radiation on the crystals of interest. 

In this work we present the results of stage (i), identifying promising crystals that justify further investigation.
The paper is arranged as follows: In section~\ref{sec:irrad} we present the crystals selected for this work and describe their irradiation with neutrons and $\gamma$s. In section~\ref{sec:optics} we describe the dedicated optical setup, fluorescence measurements, data normalization and analysis procedure. Finally, in section~\ref{sec:res} we summarize the results, and conclude in section~\ref{sec:discuss}.

\section{Irradiation campaign}
\label{sec:irrad}

The transparent monocrystallines selected for this irradiation campaign are summarized in table~\ref{tab:all_crystals}.  They were selected based on their optical transparency and known sensitivity to radiation, as well as on their commercial availability and cost to allow the realization of a future macroscopic detector, if found suitable.
The crystals are cut to a cubical shape, with at least 4 polished faces, allowing to probe the fluorescence created in the crystal bulk as opposed to the surface. No anti-reflective coating is applied, to allow measurements in a large spectral range. The crystals were not treated thermally or otherwise before the campaign with the exception of marking each face, to preserve consistent orientation in spectra measurements before and after irradiation. 
Several identical samples were used for each crystalline type. Each of the samples was irradiated once, in a single dose, as summarized in table~\ref{tab:all_crystals}.
The time between the spectral measurements and the irradiation itself was at most 5~days, both prior and post irradiation. All irradiations for this work were performed at the SOREQ Nuclear Research Center in Israel.

\begin{table}[h]
\centering
\begin{tabular}{|| l | c c c ||}
\hline
Crystal   & Supplier & Side & Samples \\
& &  (mm)& \\
\hline\hline
LiF       & UC       & 5                & 4        \\
MgF$_{2}$ & UC       & 5                & 5        \\
CaF$_{2}$ & UC       & 5                & 4        \\
BaF$_{2}$ & UC       & 5                & 4        \\
MgO       & PS       & 5                & 4        \\
Al$_{2}$O$_{3}$ & GV       & 10                & 4        \\
ZnO       & PS       & 5                & 4        \\
SiO$_{4}$ & UC       & 5                & 4        \\
LiNbO$_3$ & UC       & 5                & 3        \\
LiTaO$_3$ & UC       & 5                & 3        \\
\hline
\end{tabular}

 \caption{All monocrystallines used for this work and the number of samples used for each irradiation. Suppliers: United Crystal (UC)~\cite{uc}, Gavish Inc. (GV)~\cite{gv} and Princeton Scientific (PS)~\cite{ps}. 
 \label{tab:all_crystals}}
\end{table}

\begin{table}[t]
\centering
 \begin{tabular}{||l r | c c  ||}
 \hline
  Irradiation  & Source &   Duration &  D$_{\gamma}$ \\ 
     &  & [hour]  & [mGy]  \\ 
  
 \hline\hline 
 Neutron           &  $^{252}$Cf &    64.3     & 2.90 $\pm$ 0.75\\ 
 $\gamma_{short}$  & $^{60}$Co   &   1        &  90 $\pm$15 \\ 
 $\gamma_{medium}$ & $^{60}$Co   &   3        &  270 $\pm$41 \\ 
  $\gamma_{long}$  & $^{60}$Co   &   17.8     &  1600 $\pm$240 \\  
 \hline
 \end{tabular}
 \caption{Irradiation details. $\gamma$ doses in LiF (D$_{\gamma}$) given for reference, the $\gamma$ dose for other crystals will be proportionate to this number, depending on its absorption coefficient for $\gamma$ radiation.
}
\label{tab:all_irrad}
\end{table}

\subsection{$\gamma$ irradiation}
Crystals were exposed to a high energy $^{60}$Co $\gamma$ source for one of three distinct duration times that will be referred to as the $\gamma_{short}$, $\gamma_{medium}$ and $\gamma_{long}$ irradiations. The source activity at the time of irradiation was \SI{5.37(18)e12}{Bq}. The emitted $\gamma$ spectrum from the $^{60}$Co source contains two main energy lines at 1.17~MeV and 1.33~MeV, with a branching ratio close to 100\%, and negligible contributions from other energies. Softer photons are also expected from scattering off walls and other items.
The distance from the container holding the crystals to the source was measured to be (312~$\pm$~6)~cm. The dose at this location was measured using a calibrated TLD-100 detector, summarized in table~\ref{tab:all_irrad}. The dimensions of the crystal container allows for a maximum variation in the relative dose of 5$\%$ for different crystals. This is an upper limit for the variations, as the point-source approximation does not fully hold at this distance and the wall effects are expected to be more homogeneous. In addition, there is a 15\% systematic error in the measured $\gamma$ dose, dominated by the TLD-100 instrument error. 

The strength of the RIL generated in a specific crystal during the $^{60}$Co irradiation $S_{\rm Co}$, is modeled as a linear function of the $\gamma$ dose $D^{i}_{\gamma}$ (given in table~\ref{tab:all_irrad}) so,
\begin{equation}
\label{eq:co60}
    S_{\rm Co}^i =  \xi \cdot D^{i}_{\gamma},
\end{equation}
with the index i denoting the irradiation duration. The coefficient $\xi$ will be different for each RIL, depending on the branching ratio for its creation and its optical properties such as quantum efficiency and optical cross section.

\subsection{Neutron irradiation}
\label{sec:irrad_N}
Exposure to neutrons was performed using a $^{252}$Cf source with an activity of ($3.4\times10^8$)~Bq, providing an average of 3.7~neutron emissions per fission event.
Neutrons are emitted with energies up to 13~MeV, with an average energy of 2.13~MeV~\cite{Mannhart1987}. The crystals were placed in a lead shield and irradiated for a total duration of 64.3~hours with an estimated neutron flux of $(3.8 - 8.0)\times10^4$~n/cm$^2$/s, depending on the actual position of the crystals inside the shield.

\begin{figure}[h]
\centering

  \includegraphics[width=1\linewidth]{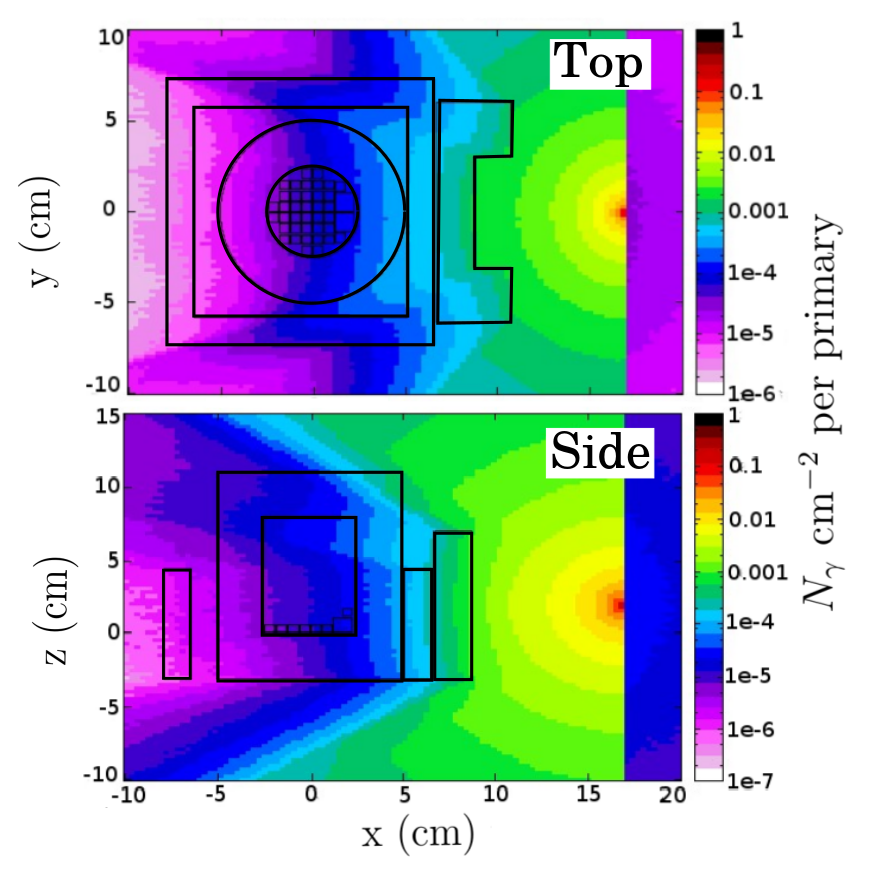}

\caption{Simulated $\gamma$s reaching the crystals per cm$^2$ per $\gamma$ primary from $^{252}$Cf source. {\it Top:} Top view, {\it Bottom:} Side view. Black lines trace the surface of Pb shield and the crystals inside the shield, with the source located at x=17.5~cm, y=0~cm, z=2.5~cm.}
\label{fig:gamma_primary_fluence_cf}
\end{figure}

\begin{figure}[h]
\centering
  \includegraphics[width=1.\linewidth]{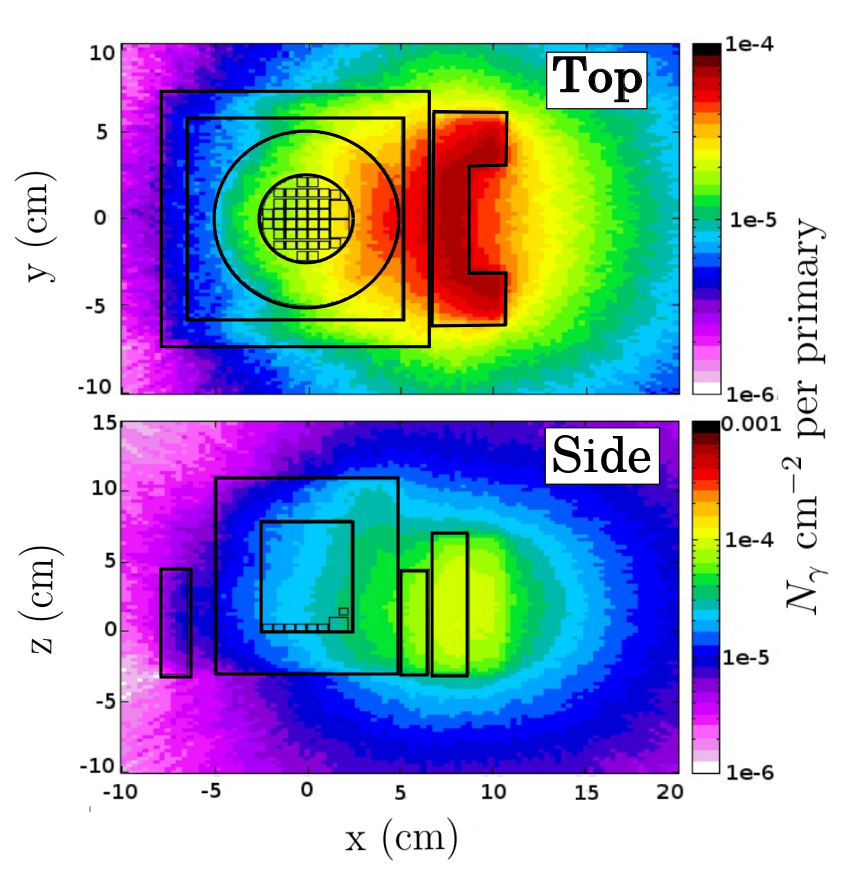}
\caption{Simulated $\gamma$s reaching the crystals, emitted from the lead shield, per cm$^2$ per neutron primary from the $^{252}$Cf source. The emitted $\gamma$s are a result of neutron activation of the lead. {\it Top:} Top view, {\it Bottom:} Side view. The geometry is identical to that of Fig.~\ref{fig:gamma_primary_fluence_cf}, with a $^{252}$Cf source at x=17.5~cm, y=0~cm, z=2.5~cm.}
\label{fig:gamma_secondary_fluence_cf}
\end{figure}

The emission of neutrons is accompanied by a significant component of $\gamma$ rays. The average number of $\gamma$s per fission event is 8.3, with an average energy of 0.8~MeV  per $\gamma$ emitted~\cite{PhysRevC.87.024601}. Lead shielding was used to reduce the flux of these $\gamma$s. However, this introduces an additional $\gamma$ source originating from the Pb activation, illustrated in Fig.~\ref{fig:gamma_secondary_fluence_cf}. The necessary thickness of the lead shield to achieve optimal n/$\gamma$ ratio was evaluated using a Monte Carlo simulation  (figs.~\ref{fig:gamma_primary_fluence_cf} and~\ref{fig:gamma_secondary_fluence_cf}) using the FLUKA software package (version 2011.2x.2)~\cite{Fluka1}, which was also used to estimate the final n/$\gamma$ ratio and $\gamma$ energy distribution in the experiment. The total $\gamma$ flux with shielding is estimated to be $(6 - 10.8)\times10^3$~$\gamma$/cm$^2$/s with an average energy of ($1.11\pm0.13$)~MeV. The fraction of $\gamma$s with energy above 2~MeV is estimated to be 10\%~-~23\%, and above 4~MeV this drops to be less than $2\%$.  Assuming that Compton scattering is the main process of energy deposition, we estimate the  $\gamma$ dose for LiF to be (2.9 $\pm$ 0.75)~mGy. 

The correct conversion factor between the effects of $\gamma$ irradiations by the $^{60}$Co and $^{252}$Cf may not be the energy deposition, and may depend on the energy spectra. This effect is estimated to increase the error of the calculated $\gamma$ dose to 50\% for the $^{252}$Cf irradiation when comparing with the $^{60}$Co doses.

The RIL in the $^{252}$Cf irradiation is modeled as originating from two distinct contributions, one from the n-radiation ($S_{n}$), the other from $\gamma$-radiation ($S_{\gamma}$). This is expressed as
\begin{equation}
 \label{eq:cf}
    S_{\rm Cf} =  S_{\gamma} + S_{n} = \xi \cdot D^{n}_{\gamma} + S_{n},
\end{equation}
where $\xi$ is defined in equation~\ref{eq:co60}, and $D^{n}_{\gamma}$ is the relative $\gamma$ dose in the neutron irradiation, taken from table~\ref{tab:all_irrad}.

The goal of this work, is to identify a clear excess of $S_n$, using the $^{60}$Co  irradiation to estimate $\xi$ and thus constrain $S_{\gamma}$. More on the statistical treatment will be presented in section~\ref{sec:analysis}

\section{Optical setup}
\label{sec:optics}
A dedicated fluorescence spectroscopy setup, tailored specifically for this campaign was designed and assembled at the Weizmann Institute. 
The optical system was designed to illuminate and collect light from within the crystal bulk, and be sensitive to a wide range of wavelengths, 250-800~nm in excitation, and 300-1200~nm for emission.
A special effort was made to use the same optical path for the entire range of excitation wavelengths, allowing for a reliable comparison of signals produced at different wavelengths. To allow a long-term campaign capable of measuring hundreds of samples under stable and repeatable conditions, the design includes \textit{in-situ} mechanical and optical calibration, real time monitoring and automation.

\subsection{Description of the setup}

\begin{figure}[h]
    \centering
    \includegraphics[width=0.9\linewidth]{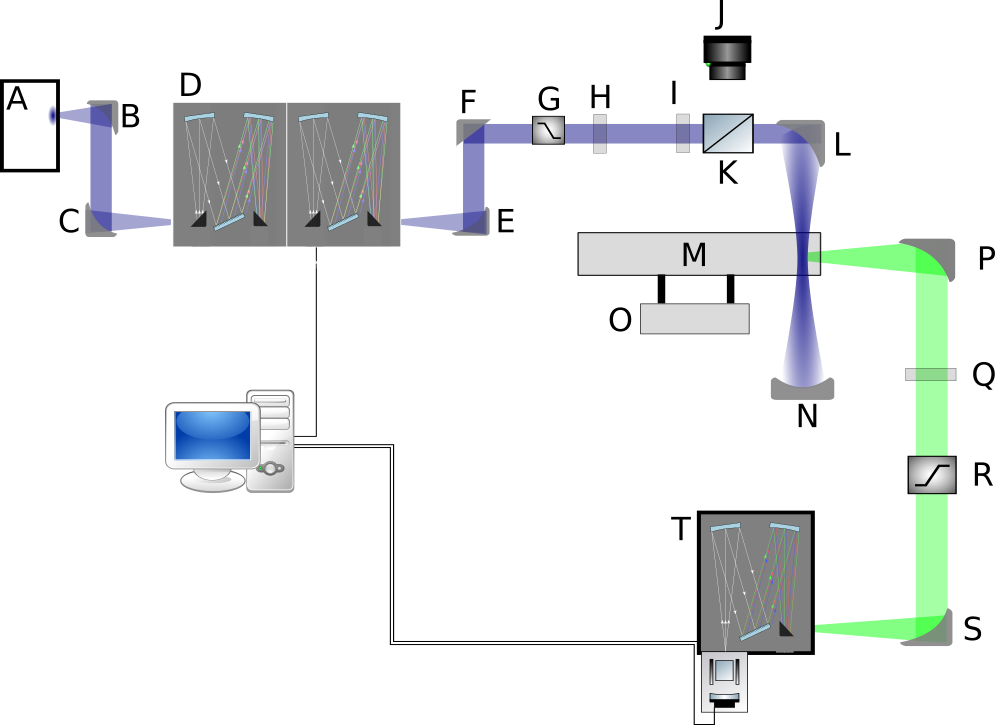}
    \caption{Schematic representation of the optical system used to measure fluorescence in this experiment. (A) Exciutation light source, (B)  Parabolic mirror, (C) Parabolic mirror, (D) Monochromator, (E) Parabolic mirror, (F) Reflective mirror, (G) Short-pass filter wheel, (H) optional long-pass filter, (I) optional polarizer, (J) Power meter, (K) Beam sampler, (L) Parabolic mirror, (M) Crystal wheel, (N) Retro-reflector (not used for the measurements reported in this paper), (O) Motorized rotation stage, (P) Parabolic mirror (Q) Optional polarizer, (R) Long-pass filter wheel, (S) Parabolic mirror, (T) Spectrograph.}
    \label{fig:fluorescence_system}
\end{figure}

A schematic of the system layout is shown in Fig.~\ref{fig:fluorescence_system}.
For optical excitation, a xenon-based, laser-excited, {\it EQ77} broadband source by {\it Energetiq} was used. Wavelength selection was performed using {\it Spectral Products CM112} Double monochromator with two gratings blazed at 300 and 500~nm, both with a line density of 1200 mm$^{-1}$, giving a band-pass of approximately 5~nm.
Thorlabs edgepass short pass filters were used on the excitation side in steps of 50~nm from 400~nm to 600~nm cutoffs, to reduce stray light from the monochromator that overlaps with the wavelength range being measured. Long pass filters were used in steps of 50~nm from 450~nm to 650~nm cutoff to reduce the scattered excitation light from entering the spectrograph. An additional long pass filter with a cutoff at 355~nm was placed in the excitation beam to clean out second harmonics from the monochromator when exciting with wavelengths longer than 360~nm. 

Emission spectra were measured using the {\it Shemrock 193i} spectrograph with a {\it Andor Newton 971 camera}. Motorized stages were used to automate the filter wheels, crystal wheel, excitation focusing mirror and emission collection mirror. 

To account for fluctuations of the light source power, 
A beam sampler is permanently placed in the excitation beam path with a photodiode perpendicular to the beam.

The crystal holder allows 12~crystals to be loaded simultaneously. Four positions were reserved for permanent references, three reference crystals with known signal, and one empty crystal holder. These references were measured following every loading of new crystals.  These purpose of these auxiliary measurements was to monitor and correct for changes in background or efficiency between measurements.
 
 \begin{figure}[h]
     \centering
     \includegraphics[width=0.95\linewidth]{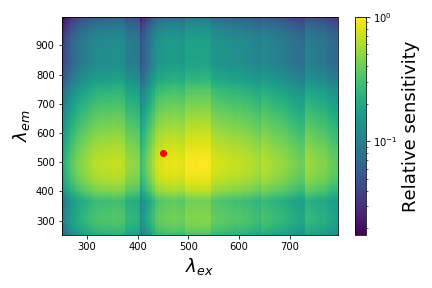}
     \caption{Measured relative detection sensitivity, in log scale, of the optical system as a function of excitation and emission wavelengths $\lambda_{ex}$, $\lambda_{em}$. The visible discontinuities are mostly due to changing optical filters, except for the ones at $\lambda_{ex}$ and $\lambda_{ex}$ of 400~nm, which are due to changing monochromator and spectrograph gratings respectively. The red circle marks the point that was measured in absolute value using the 3410 cm$^{-1}$ Raman line of water.}
     \label{fig:rel_corr}
 \end{figure}

\subsection{Estimation of the experiment sensitivity}
 The sensitivity to detect a given active CC is estimated, in order to assess the meaning of null results as well as future comparisons with simulations. Null results can be translated to an upper limit on  branching ratio of production, with an unknown pre-factor of the optical excitation cross section.
The expected sensitivity was estimated using the Raman line of water at 3410 cm$^{-1}$. This technique is commonly used in absolute spectrofluorometry to present absolute cross sections~\cite{BOCKLITZ2015544,lawaetz_fluorescence_2009}. The measurement is given here as the minimum detectable defect density, for a defect with a cross section of 10$^{-17}$~cm$^2$ and unity quantum efficiency, emitting non-polarized light, and a detection threshold of 10 photo-electrons per second. Signals with a different cross section or in a crystal with a different refractive index, will give different values. The sensitivity presented here allows only rough limits on the density of created CCs. We estimate a sensitivity of
\begin{equation}
    \rho_{min} = 3.7(7) \cdot 10^5 \mathrm{cm}^{-3} \frac{10^{-17} \mathrm{cm}^2}{ QE \cdot \sigma_{optical}}
\end{equation}
at excitation wavelength of 450~nm and emission at 532~nm (illustrated in Fig. \ref{fig:rel_corr}), where $\sigma_{optical}$ is the optical cross section of the defect creating the signal and $QE$ is the defects quantum efficiency.
For detected signals, this experiment allows us to estimate the capabilities of a hypothetical detector for the detection of pure nuclear interactions. The details of how many color centers of each kind are created and their respective properties, such as optical cross section and quantum efficiency, can be ignored if the crystal response scales linearly with the the rate of recoils and is homogeneous over the bulk of the crystal. The deviation from these assumption will be the subject of further study. Assuming the response is linear at low rates, the hypothetical signal in a detector can be inferred for a given rate of recoils, as a function of the optical collection efficiency achieved, for each of the detected signals. 

\subsection{Crystal measurements}
Each crystal was excited with  wavelengths between 250~nm and 800~nm in steps of 10~nm, scanning down from longer wavelength towards shorter ones to reduce the risk of signal bleaching by UV light prior to its measurement. Each emission spectrum was measured multiple times with varying CCD exposure times T$_{CCD}$. An exposure of 1~s was repeated 4 times to identify bleaching during the measurement. Additional exposures of 0.1~s and 10~s were performed to widen dynamic range.
While scanning the crystals of interest, a scan through the auxiliary measurements is also performed. 

\begin{figure}[h]
\centering

  \includegraphics[width=1.\linewidth]{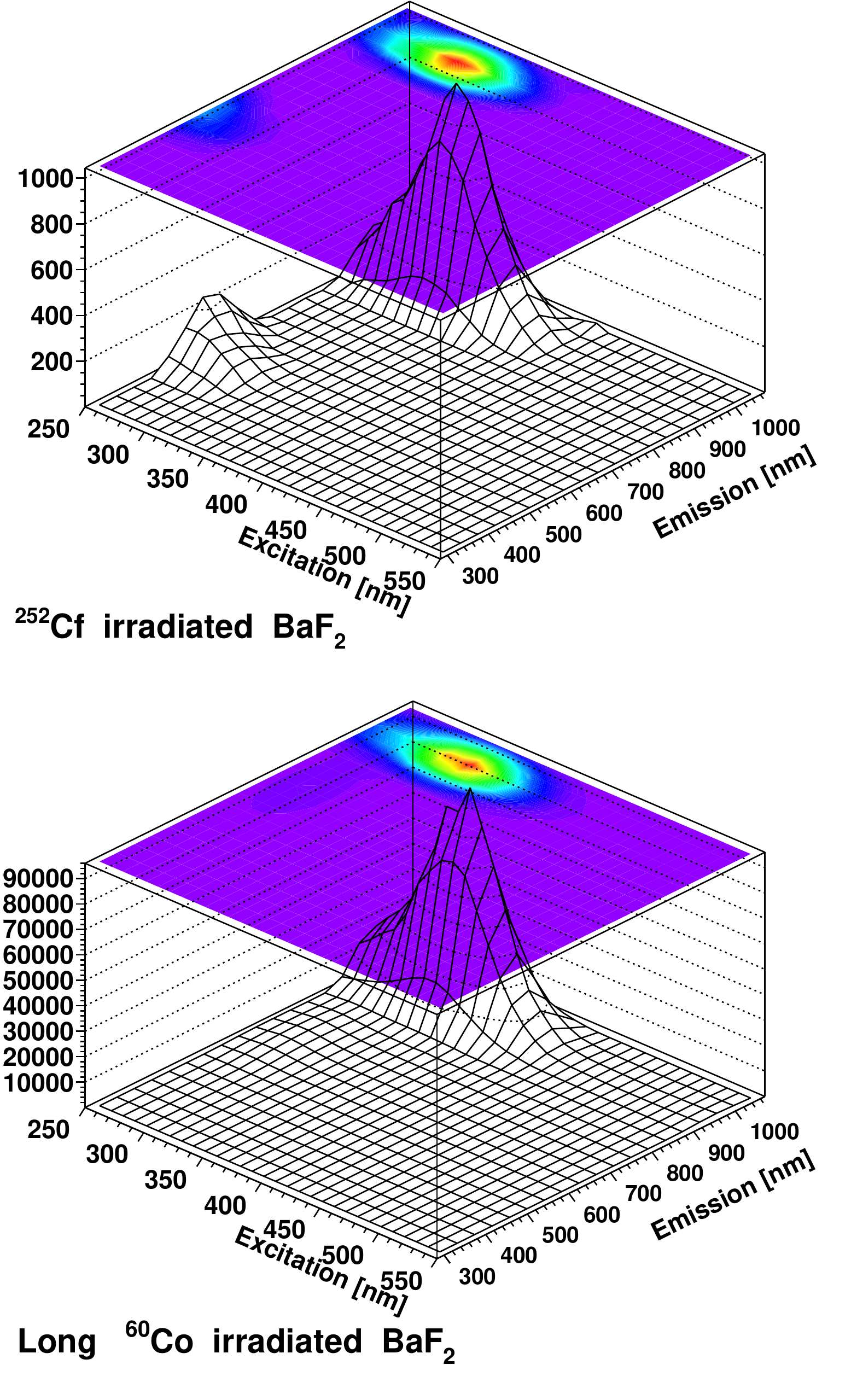}

\caption{RIL spectra for two BaF$_2$ crystal. One irradiated with the $^{252}$Cf source (\textit{top}), and one with the long duration $^{60}$Co (\textit{bottom}). Notice the 570~nm emission, excited at 340~nm RIL signal that appears only in the $^{252}$Cf irradiation.
}

\label{fig:2D_scan}
\end{figure}

The measured data was normalized to counts per excitation photon per second. Before normalization, the measured background (signal baseline) N$_{BG}(\lambda_{em})$ was subtracted from the measured quanta N$_{meas}$. The measured spectra was normalized by the relative number of excitation photons N$_{ex}(\lambda_{ex})$, the CCD exposure time $T_{CCD}$, and the relative collection efficiency of the system expressed as $\eta$. The final spectra used for analysis N$_{norm}(\lambda_{ex},\lambda_{em})$ can be expressed as

\begin{equation}
 {\rm N}_{norm}(\lambda_{ex},\lambda_{em}) = \frac{{\rm N}_{meas}- {\rm N}_{BG}(\lambda_{em})}{\eta(\lambda_{em})\times {\rm N}_{ex}(\lambda_{ex})\times T_{CCD}},
\end{equation}
which gives the normalized detected quanta per excitation quanta.
Since each measurement was performed multiple times with different  exposures times, the measurement with the longest non-saturated $T_{CCD}$  was  used. All analysis results presented here were performed on the increase of signal due to irradiation, i.e. the  pre-irradiation spectra were subtracted from the post-irradiation ones (see Fig.~\ref{fig:2D_scan} top).

\subsection{Analysis}
\label{sec:analysis}
\begin{figure}[h]
\centering
  \includegraphics[width=0.95\linewidth]{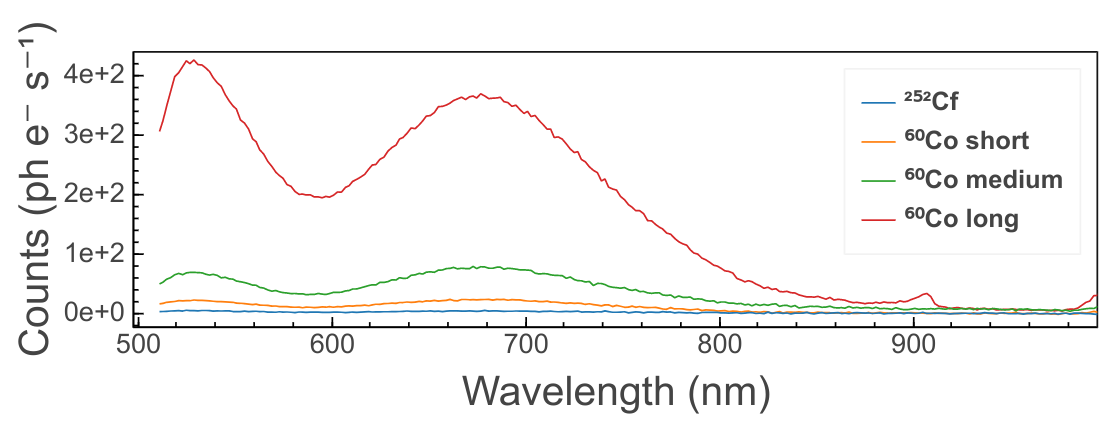}
    \includegraphics[width=0.95\linewidth]{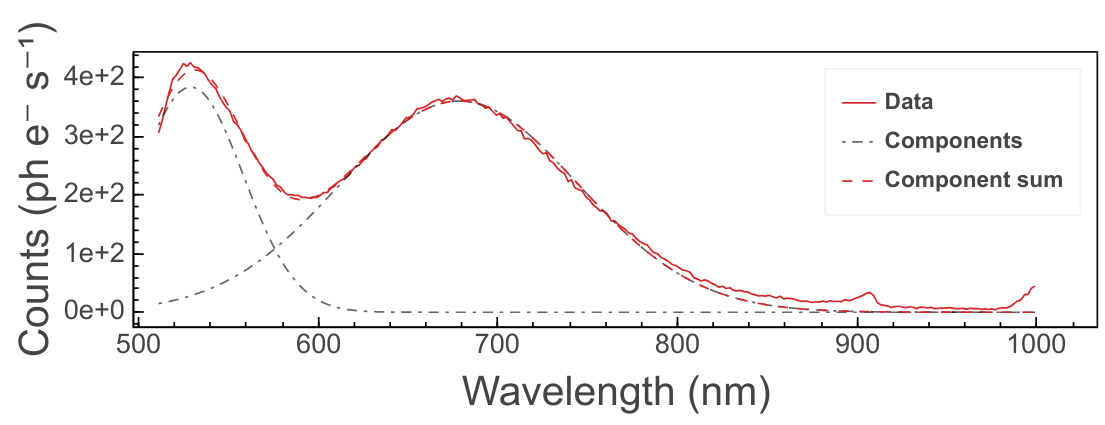}
\caption{The LiF spectra, when excited by 450~nm light. Example showing spectra of the signal excess created by the four irradiations (\textit{Top}), and the spectral deconvolution into individual components by fitting a sum of Gaussians (\textit{Bottom}).}
\label{fig:spectra_plots}
\end{figure}

 Spectrum deconvolution was done by fitting to a sum of Gaussian models, since the signal width appears to be dominated by the inhomogeneous broadening in all cases. The total signal from each peak was then estimated as the area under the best fit Gaussian at the excitation wavelength with maximum signal, this is illustrated for LiF in Fig.~\ref{fig:spectra_plots} bottom.
 
For signals that are identified across all 4 irradiations (i.e. $S_{\rm Cf}$, $S_{\rm Co}^{short}$, $S_{\rm Co}^{medium}$, $S_{\rm Co}^{long}$), a z-score and confidence interval for S$_{n}$ was calculated using a $\chi^{2}$ test statistic for hypothesis testing. The test statistic was used to reject the null hypothesis (corresponding to S$_{n}$=0 in eq.~\ref{eq:cf}), for the alternative hypothesis with an additional signal S$_{n}$=\^{S}$_{n}$  created by the neutrons in the $^{252}$Cf irradiation. Where \^{S}$_{n}$ is the value of S$_{n}$ that minimizes $\chi^{2}$ .

\begin{figure}[h]
\centering

   \includegraphics[width=0.95\linewidth]{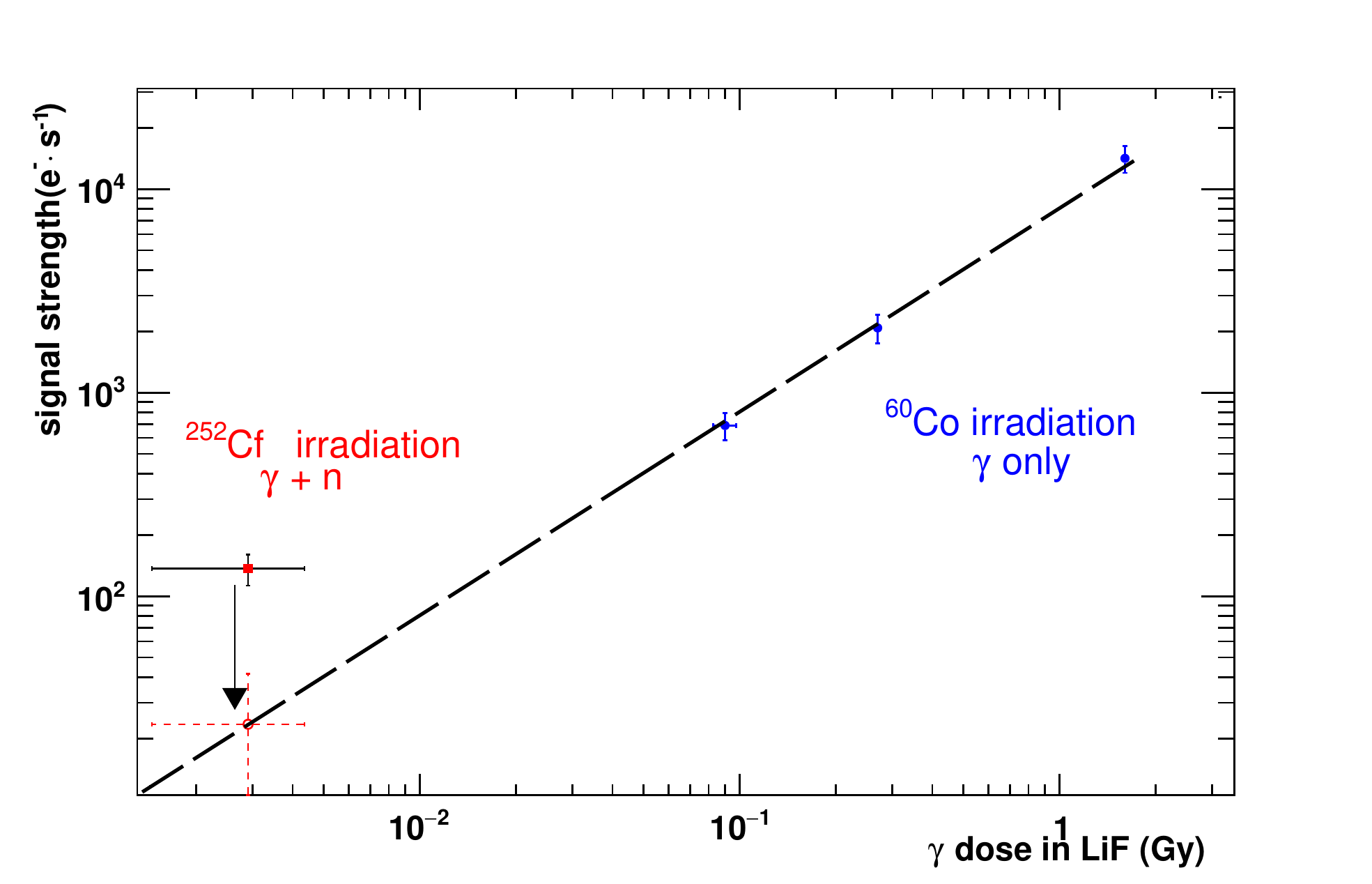}
\caption{Example of a fit to signal strength as a function of relative  $\gamma$ dose for the model of linear dependence on the $\gamma$ dose plus an unknown contribution from the neutrons in the $^{252}$Cf irradiation. Shown here for LiF at 530~nm emission, excited at 450~nm. The RIL from the $^{60}$Co irradiations is shown as blue circles, measured RIL from the $^{252}$Cf shown as red circle, expected RIL from the $\gamma$s present during the irradiation shown as red dashed cross and linear fit $^{60}$Co irradiations as black dashed line.}
\label{fig:model_fit_lif_450_530}
\end{figure}

\begin{figure}[h]
\centering
  \includegraphics[width=0.95\linewidth]{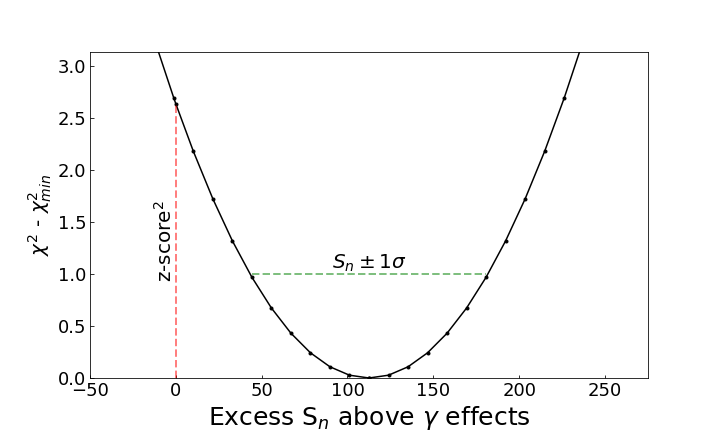}
    
\caption{Test statistic as a function of excess S$_{n}$ above $\gamma$ effects. Assuming asymptotic distribution, the confidence interval is taken as the all the points with a $\chi^{2}$ smaller than the minimum $\chi^2$ + 1. The significance is equal to the square root of the difference in the test statistic between its lowest value and its value where S$_{n}$=0. Example shown here from LiF at 530~nm emission, excited at 450~nm.}
\label{fig:liklihood_lif_450_530}
\end{figure}

An example fit is shown in Fig.~\ref{fig:model_fit_lif_450_530} for LiF.  When calculating the significance and z-score, we assume the test statistic is asymptotically distributed, an illustration of this is shown in Fig.~\ref{fig:liklihood_lif_450_530}.

\section{Results}
\label{sec:res}
We present the results of the irradiations and measurements, after the analysis procedures described above. First we show the signals that have been enhanced by the various irradiations, and then discuss the bleaching of these signals.
\subsection{Signals produced in irradiations}
\label{subsec:sigs}
\begin{figure}
\centering
    \includegraphics[width=0.95\linewidth]{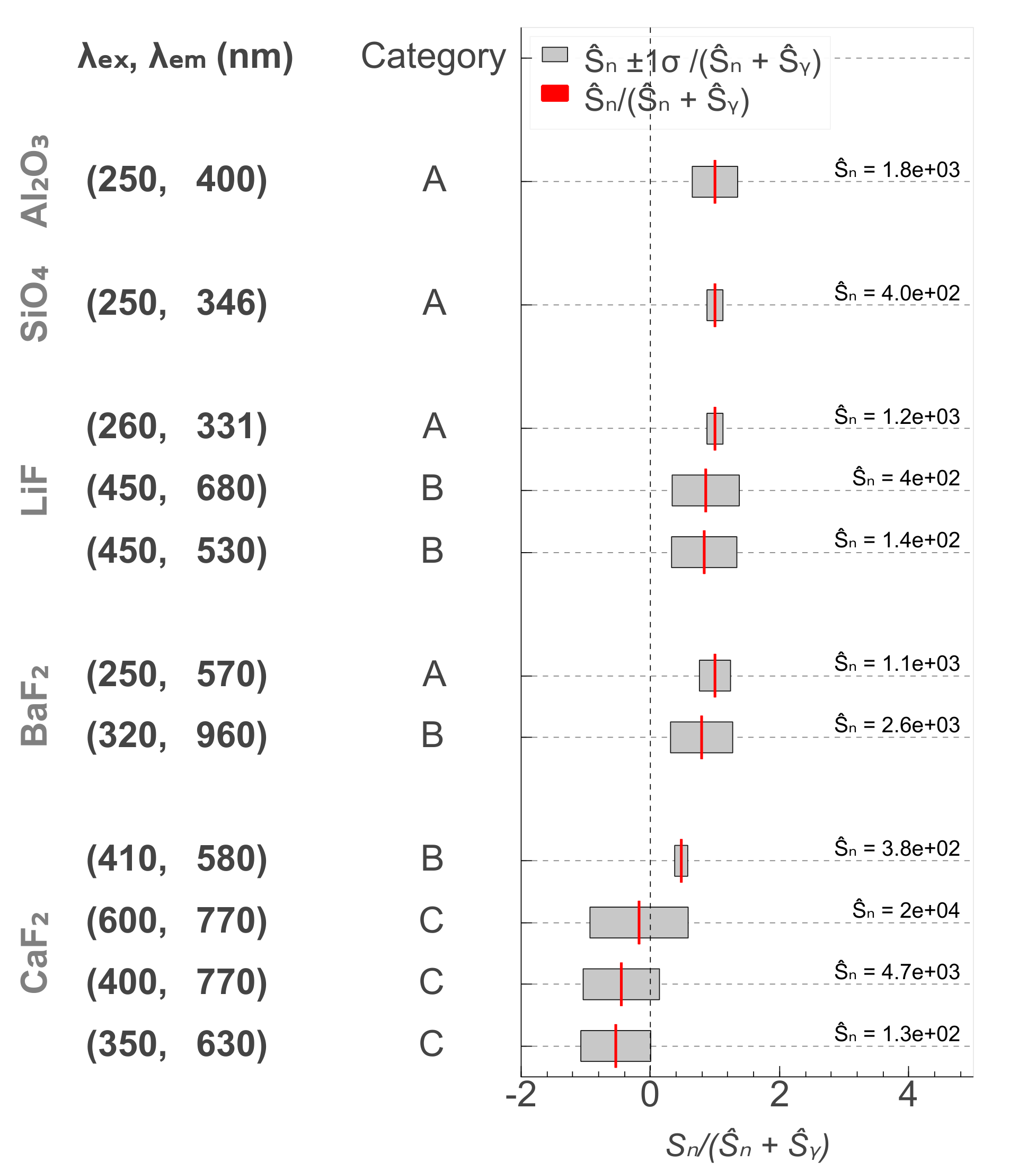}
\caption{Best fit values (in red) and confidence intervals (grey) for the RIL attributed to the neutron radiation as opposed to the $\gamma$ rays in the $^{252}$Cf irradiation. Categories defined in section \ref{sec:res}. }

\label{fig:excess_plot}
\end{figure}

The RIL observed in the various measured crystals, can be divided into a number of distinct categories listed by their importance to this study:
\begin{description}

\item[Category A] RIL observed only in the $^{252}$Cf irradiation and in none of the $^{60}$Co irradiations. The luminescence is therefore attributed solely to neutrons. Shown in Fig.~\ref{fig:excess_plot}.

\item[Category B] RIL observed for all irradiations, for which a neutron induced signal was detected with a significance above 1~$\sigma$. Shown in Fig.~\ref{fig:excess_plot}.

\item[Category C] RIL observed for all irradiations, consistent with S$_{n}$=0 (no detected neutron induced signal). Shown in Fig.~\ref{fig:excess_plot}.

\item[Category D] Observed nonlinear ratio between the RIL for the different $^{60}$Co doses. This can be either due to strong orientation dependence, polarization dependence or nonlinearity in the creation mechanism itself. Shown in table~\ref{tab:qualitative_table}.

\item[Category E] RIL observed only for the largest dose $^{60}$Co irradiation, which is beyond the interest of this paper. Shown in table~\ref{tab:qualitative_table}.

\end{description}

In addition to these clearly identified RILs, the LiNbO$_3$ \& LiTaO$_3$ crystals gave null results, in which no RIL was observed and MgO included too much existing signal to distinguish clear RIL and therefore was not considered interesting for this study.

\begin{table}[t]

\centering 
 \begin{tabular}{|| l | l l | l |} 
 \hline
    Crystal  & $\lambda_{ex}$ & $\lambda_{em}$  & Category  \\
    \hline	
    CaF$_{2}$& 340 (70) & 425 (50) & D \\
             & 520 (70)	& 635 (70) & E \\
             
    \hline
    MgF$_{2}$& 250 (20) & 420 (50)& D \\
             & 420 (35)	& 560 (90) & D \\
             & 420 (50)	& 1000 (200) & D \\
             & 510 (55)	& 670 (100) & D \\
    \hline
    ZnO & 350 (140) & 580 (140) & D \\
    
\hline

\end{tabular}
\caption{List of observed signals induced by irradiation that do not meet the criteria for quantitative analysis in this study.}
\label{tab:qualitative_table}
\end{table}

\subsection{Bleaching of signals}
Using the consecutive irradiations performed at each excitation wavelength, we estimate the bleaching per 10~s of exposure to the excitation at the wavelength corresponding to the peak of the signal. We have observed bleaching in BaF at (320, 960)~nm, CaF (440, 700)~nm, MgF (370, 600)~nm, of $2.2(7)\%$, $5.8(2)\%$, $10.5(5)\%$ respectively per 10 seconds of measurement. Future studies will look into bleaching of centers of interest with higher accuracy.

\section{Discussion}
\label{sec:discuss}
This concludes the first stage in assessing the viability of various transparent crystals to serve as detectors for low mass dark matter. We aim to contrast the production of CCs by fast neutrons with those produced by $\gamma$ radiation. Both the production by $\gamma$ rays and neutrons are of interest, for purposes of background and signal characterization, respectively.  Focusing on commercially available single crystals we map a number of RILs that justify further investigation.
The results are concisely summarized in Fig.~\ref{fig:excess_plot} and table~\ref{tab:qualitative_table}. The production rates by $\gamma$ radiation can be used to study desired background levels for various crystals and CCs. For the neutron irradiation, we focus on a number of defects that have shown evidence of excess over $\gamma$ effects. The CCs we have found in that category are in the  BaF$_2$, CaF$_2$, SiO$_4$, Al$_2$O$_3$ and LiF crystals.
These CCs will be studied further with higher priority as they show indications of a creation mechanism involving nuclear recoils. We draw the readers attention to the fact that some of these CCs are at the edge of our range in excitation wavelength and therefore need to be verified in our next study after upgrading our system to operate at wavelengths below 250~nm. It should be mentioned that the rate of $\gamma$ irradiation used in this study is orders of magnitude higher than that expected in a rare-event environment. Later studies will be required to verify that the defects induced by the $\gamma$ source repeat with much lower activities, and are not a result of nonlinear effects, which would not exist in the low background setting of a DM experiment. This assumption is supported in most CCs studied here in which a linear dependence fit very well to data.
This paper is the first experimental step in a long campaign, which will be followed by studies dedicated to the search and characterization of crystal defects as low mass DM detectors. The response to lower energy neutrons, the annealing of defects, optical bleaching and stability will be the subjects of later studies. In addition to further investigating the crystals reported here, additional candidate crystals may be investigated in the future. The goal being identification of a viable candidate for such a detector and the construction of a prototype detector.

Acknowledgements: 
Many people have assisted along the way and we would like to express our gratitude, mostly to A. Manfredini, A. Marin, T. Volansky, O. Slone, A. Soffer, D. Leppert-Simenauer, I. Sagiv, M. Shutman and M. M. Devi. We are particularly grateful for the assistance given by O. Heber with the earlier measurements. This work was supported by the PAZY foundation  and  by the Minerva foundation with funding from the Federal German Ministry for Education and Research.
RB is the incumbent of the Arye and Ido Dissentshik Career Development Chair.

\bibliographystyle{elsarticle-num}
\bibliography{CC.bib}

\end{document}